\documentclass[journal]{IEEEtran}
\usepackage{cite}

\ifCLASSINFOpdf
  \usepackage[pdftex]{graphicx}
\else
\fi
\usepackage{amsmath}
\usepackage{amssymb}
\usepackage{bm}
\usepackage{bbm}
\usepackage{algorithmic}
\usepackage{xcolor}

\begin{document}
%
\title{Design of Tight Minimum-Sidelobe Windows \\ by Riemannian Newton's Method}
%
%
%

\author{Daichi~Kitahara,~\IEEEmembership{Member,~IEEE,}
        and~Kohei~Yatabe,~\IEEEmembership{Member,~IEEE}
\thanks{D. Kitahara and K. Yatabe contributed equally to this work.}
\thanks{D. Kitahara is with Ritsumeikan University.}
\thanks{K. Yatabe is with Waseda University.}
\thanks{Manuscript received XXXX XX, 2021; revised XXXX XX, 2021.}}

%
%

\markboth{Signal Processing Letters,~Vol.~XX, No.~X, XXXX~202X}%
{Kitahara \& Yatabe: Design of Tight Minimum-Sidelobe Windows by Riemannian Newton's Method}
%



\maketitle

\begin{abstract}
The short-time Fourier transform (STFT), or the discrete Gabor transform (DGT), has been extensively used in signal analysis and processing.
Their properties are characterized by a window function.
For signal processing, designing a special window called \textit{tight window} is important because it is known to make DGT-domain processing robust to error.
In this paper, we propose a method of designing tight windows that minimize the sidelobe energy.
It is formulated as a constrained spectral concentration problem, and a Newton's method on an oblique manifold is derived to efficiently obtain a solution.
Our numerical example showed that the proposed algorithm requires only several iterations to reach a stationary point.
\end{abstract}


%
\IEEEpeerreviewmaketitle

\section{Introduction}
%
%
%
%
\IEEEPARstart{T}{he} short-time Fourier transform (STFT), or the discrete Gabor transform (DGT) \cite{GaborAlgorithm1998,Foundation2001,TFconcepts2008,framesBases,TenLectures1992}, has been extensively used as the standard tool for signal analysis and processing \cite{Kitahara2020a,Kitahara2020b,atomic,HVA,RBM,DeGLI,ConsistILRMA,SpecResDNN,phase}.
The major advantage of STFT/DGT over the Fourier transform is its ability to localize the frequency components of a signal at each time.
This is realized by a tapering function called \textit{window} whose energy is concentrated in the time-frequency domain.
Since the properties of STFT/DGT are characterized by the used window, many windows have been proposed in the literature \cite{HarrisProcIEEE1978,Slepian1978,Kaiser1980,NuttallWin1981,Rohling1983,Prabhu1989,Bspline1989,Kulkarni2003,Bergen2004,Desbiens2006,Kawahara2017,Kusano2020}.

While most of the existing windows have been designed for signal \textit{analysis}, windows for \textit{synthesis} are equally important for processing.
STFT/DGT-domain signal processing modifies STFT/DGT coefficients and then converts it back to the time domain.
In the forward and inverse transformations, the signal is multiplied by analysis and synthesis windows, respectively.
Therefore, both analysis and synthesis windows affect the result of STFT/DGT-domain processing.

The common practice is to firstly choose a window that is optimized for analysis, and then a synthesis window is generated to satisfy the perfect reconstruction condition.
Such a window that perfectly reconstructs an unprocessed signal is said to be \textit{dual} \cite{GaborAlgorithm1998,Foundation2001,TFconcepts2008,framesBases,TenLectures1992}.
The most used dual window for synthesis is the canonical, or minimum-norm, dual window, and hence its computation is well-studied \cite{CanonDual1995,Janssen2002,Janssen2007,Sondergaard2012}.
However, such canonical windows are not optimized for processing, i.e., there must exist a better dual window suitable for signal processing.
Therefore, methods for designing other types of dual windows have also been proposed \cite{DualFrame2005,ConvexOpt2018}.

An important class of windows for signal processing is \textit{tight window}%
\footnote{To be accurate, a tight window is a window that generates a tight Gabor frame \cite{GaborAlgorithm1998,Foundation2001,TFconcepts2008,framesBases,TenLectures1992}. We use the term ``tight window'' throughout the paper because our focus is on a window and not on the associated frame.} \cite{Janssen2002}.
A tight window can perfectly reconstruct a signal by using it for both analysis and synthesis, i.e., it is self-dual.
According to the frame theory,
tight windows can make DGT robust to error of processing \cite{Life1,Life2}.
Therefore, their design should be an important topic for DGT-domain processing.
However, design of tight windows has been rarely studied until recently \cite{ConvexOpt2018,ICASSP2019DPSS,ICA2019NearlyTight} probably due to the complicated nature of the set of tight windows.

In this paper, we propose a method of designing tight windows that minimize the sidelobe energy.
Since a set of tight windows can be represented as an oblique manifold, we formulate the window design as a manifold optimization problem.
Then, we derive a Riemannian Newton's method that can compute an optimal tight window by several iterations.

\section{Preliminaries}
Let $\mathbb R$, $\mathbb C$, and $\mathbb N$ be the sets of all real numbers, all complex numbers, and all nonnegative integers, respectively.
The imaginary unit is denoted by $\mathbbm{i}\in\mathbb C$, i.e., $\mathbbm{i}^{2}=-1$.
For any complex number $c\in\mathbb C$, $\bar{c}$ and $|c|:=\sqrt{c\bar{c}}$ denote the complex conjugate and the absolute value of $c$.
We write vectors and matrices by boldface lowercase and uppercase letters, respectively, and their transposes are denoted by $(\cdot)^{\mathrm{T}}$.
The identity matrix of order $N$ is denoted by $\bm{I}_{N}\in\{0,1\}^{N\times N}$, and a zero matrix of size $M\times N$ is denoted by $\bm{O}_{M\times N}\in\{0\}^{M\times N}$.
For any real vectors $\bm{x},\bm{y}\in\mathbb R^{N}$, the Euclidean metric, or the standard inner product, is defined as $\langle \bm{x},\bm{y} \rangle=\bm{x}^{\mathrm{T}}\bm{y}$, and the Euclidean norm of $\bm{x}$ is defined as $\lVert\bm{x}\rVert_{2}=\sqrt{\langle \bm{x},\bm{x} \rangle}=\sqrt{\bm{x}^{\mathrm{T}}\bm{x}}$.
A sphere of radius $r$ in $\mathbb R^{N}$ is defined as $\mathcal S^{N-1}_{r}=\{\bm{x}\in\mathbb R^{N}\,|\,\lVert \bm{x}\rVert_{2}=r\}$.
The floor and ceiling functions are denoted by $\lfloor \cdot \rfloor$ and $\lceil \cdot \rceil$, respectively.
The order of a function is denoted by $\mathcal O(\cdot)$.

\subsection{Discrete Gabor Transform and Its Inversion}
Let $\bm{x}:=(x[0],x[1],\ldots,x[L-1])^{\mathrm T}\in\mathbb C^{L}$ be a discrete-time signal of length $L$, and $\bm{w}:=(w[0],w[1],\ldots,w[K-1])^{\mathrm T}\in\mathbb C^{K}$ be a window of length $K$ such that $\bm{w}$ is shorter than $\bm{x}$, i.e., $K<L$.
In addition, let $a$ and $M$ be integers satisfying 
$L/a=:N\in\mathbb N$ and $0<a<K\leq M \leq L$.
In this paper, we define STFT/DGT of $\bm{x}$ with respect to the window $\bm{w}$ as
\begin{equation}
X[m,n]=\sum_{l=0}^{K-1}x[l+an]\,\overline{w[l]}\, \mathrm{e}^{-2\pi \mathbbm{i}m l/M}\mbox{,}
\label{DGT}
\end{equation}
where $m=0,1,\ldots,M-1$, $n=0,1,\ldots,N-1$, and the signal $x[\cdot]$ is treated periodically as $x[l+L]:=x[l]$.

The \textit{inverse DGT} is defined with a synthesis window $\bm{\gamma}$ as
\begin{equation}
x[l]=\sum_{n=\lceil(l-K+1)/a\rceil}^{\lfloor l/a\rfloor }\gamma[l-an]\sum_{m=0}^{M-1}X[m,n]\, \mathrm{e}^{2\pi \mathbbm{i}m (l-an)/M}\mbox{,}
\label{IDGT}
\end{equation}
where $X[m,-n]$ is treated as $X[m,-n]:=X[m,N-n]$ for $n=1,2,\ldots,\lfloor(K-1)/a\rfloor$.
This transform reconstructs the original signal if $\bm{\gamma}$ is dual of $\bm{w}$ \cite{Foundation2001}.
When a window $\bm{w}$ is self-dual (i.e., signals can be reconstructed by using $\bm{w}/\lambda$ in place of $\bm{\gamma}$ with some $\lambda>0$), it is called a tight window.
A tight window $\bm{w}_\text{t}$ can be obtained by the following projection:
\begin{equation}
\bm{w}_\text{t}=\sqrt{\lambda}\bm{S}^{-1/2}_{\bm{w}}\bm{w}\mbox{,}\label{eq:tightWin}
\end{equation}
where $\bm{S}_{\bm{w}}\in\mathbb R^{K\times K}$ is a diagonal matrix corresponding to the \textit{frame operator} \cite{framesBases}, and its diagonal components are given by
\begin{equation}
[\bm{S}_{\bm{w}}]_{l,l}=M\sum_{n=-\lfloor l/a\rfloor}^{\lfloor(K-l-1)/a\rfloor}|w[l+an]|^2\quad\mbox{(}l=0,1,\ldots,K-1\mbox{).}\label{frame_operator}
\end{equation}
When $\lambda=1$, the tight window $\bm{w}_\text{t}$ obtained by \eqref{eq:tightWin} is called the \textit{canonical tight window} \cite{Janssen2002}.
In this paper, we assume that the window $\bm{w}$ is real-valued for simplicity, i.e., $\bm{w},\bm{\gamma}\in\mathbb R^{K}$.

\subsection{Newton's Method on Riemannian Manifold}
Let $\mathcal M$ ($\subset \mathbb R^{N}$) be a \textit{Riemannian manifold} embedded in the Euclidean space,%
\footnote{$\mathcal M$ is also called a \textit{Riemannian submanifold of $\mathbb R^{N}$ with the metric $\langle \cdot,\cdot \rangle$}.}
where the \textit{tangent space to $\mathcal M$ at $\bm{x}\in\mathcal M$} is denoted by $T_{\bm{x}}{\mathcal M}$.
Let $h:\mathbb R^{N}\to\mathbb R$ be a twice continuously differentiable function whose gradient in the Euclidean metric is $\nabla h:\mathbb R^{N}\to\mathbb R^{N}$.
We address the following problem on $\mathcal M$:
\begin{equation}
\mathop{\mathrm{minimize}}_{\bm{x}\,\in\, \mathcal M}\,h(\bm{x})\mbox{.}\label{opt_on_M}
\end{equation}

To solve the problem in \eqref{opt_on_M}, we introduce the Riemannian gradient and Hessian of $h$ on $\mathcal M$ \cite{Absil-Mahony-Sepulchre-07,boumal2020intromanifolds,Hu-Liu-Wen-Yuan-20,Sato-21}.
The \textit{Riemannian gradient at $\bm{x}\in\mathcal M$} is a vector $\bm{g}_{\bm{x}}\in T_{\bm{x}}\mathcal M$ ($\subset \mathbb R^{N}$) defined~as
\begin{equation}
\bm{g}_{\bm{x}}=(g_{1}(\bm{x}),g_{2}(\bm{x}),\ldots,g_{N}(\bm{x}))^{\mathrm T}=\bm{P}_{\bm{x}}\bigl(\nabla h(\bm{x})\bigr)\mbox{,}\label{gradient}
\end{equation}
where $g_{n}:\mathbb R^{N}\to \mathbb R$ ($n=1,2,\ldots,N$) is a differentiable function (a component of $\bm{g}_{\bm{x}}$), and $\bm{P}_{\bm{x}}:\mathbb R^{N}\to T_{\bm{x}}\mathcal M$ is the \textit{metric projection onto $T_{\bm{x}}\mathcal M$}.
This projection is given as an orthogonal projection matrix $\bm{P}_{\bm{x}}\in\mathbb R^{N\times N}$ since $T_{\bm{x}}\mathcal M$ is a linear subspace for any $\bm{x}\in\mathcal M$.
Next, the \textit{Riemannian Hessian at $\bm{x} \in\mathcal M$} is defined as a linear operator $\bm{H}_{\bm{x}}:T_{\bm{x}}\mathcal M\to T_{\bm{x}}\mathcal M$ and can be expressed as some matrix $\bm{H}_{\bm{x}}\in\mathbb R^{N\times N}$ satisfying%
\footnote{$\bm{H}_{\bm{x}}=\bm{P}_{\bm{x}}\,[\nabla g_{1}(\bm{x}) \ \ \nabla g_{2}(\bm{x}) \ \cdots\ \nabla g_{N}(\bm{x})]^{\mathrm{T}}$ does not have to hold in \eqref{Hessian} since $\bm{v}$ belongs to the tangent space $T_{\bm{x}}\mathcal M$, not the Euclidean space $\mathbb R^{N}$.}
\begin{equation}
\bm{H}_{\bm{x}}\bm{v}=\bm{P}_{\bm{x}}\Bigl(
\begin{bmatrix}
\nabla g_{1}(\bm{x}) & \nabla g_{2}(\bm{x}) & \cdots&\nabla g_{N}(\bm{x})
\end{bmatrix}^{\mathrm{T}}
\bm{v}\Bigr)\label{Hessian}
\end{equation}
for all $\bm{v}\in T_{\bm{x}}\mathcal M$.
We also define a mapping $R_{\bm{x}}:T_{\bm{x}}\mathcal M\to\mathcal M$ as $R_{\bm{x}}(\bm{v})= P_{\mathcal M}(\bm{x}+\bm{v})$, where $P_{\mathcal M}:\mathbb R^{N}\to \mathcal M$ is the single-valued metric projection onto $\mathcal M$.
From \cite[Propositions~5.43, 5.47, and 5.48]{boumal2020intromanifolds}, $R_{\bm{x}}$ is a \textit{second-order retraction on $\mathcal M$}, and
\begin{equation}
h(R_{\bm{x}}(\bm{v})) = h(\bm{x})+\langle \bm{g}_{\bm{x}},\bm{v}\rangle +\frac{1}{2}\langle \bm{H}_{\bm{x}}\bm{v},\bm{v}\rangle+\mathcal O(\lVert \bm{v}\rVert^{3}_{2})\label{quad_approx}
\end{equation}
holds for all $\bm{x}\in \mathcal M$ and all $\bm{v}\in T_{\bm{x}}\mathcal M$.
If the inverse operator $\bm{H}^{-1}_{\bm{x}}:T_{\bm{x}}\mathcal M\to T_{\bm{x}}\mathcal M$ of $\bm{H}_{\bm{x}}$ exists, a stationary point $\bm{v}\in T_{\bm{x}}\mathcal M$ of the function $h\circ R_{\bm{x}}:T_{\bm{x}}\mathcal M\to\mathbb R$ is expressed as $\bm{v}=-\bm{H}^{-1}_{\bm{x}}\bm{g}_{\bm{x}}$ by ignoring the last term of \eqref{quad_approx}.

The \textit{Riemannian Newton's method} solves \eqref{opt_on_M} by iterating
\begin{equation}
\bm{x}^{(i+1)}=R_{\bm{x}^{(i)}}\bigl(-\bm{H}_{\bm{x}^{(i)}}^{-1}\bm{g}_{\bm{x}^{(i)}}\bigr)= P_{\mathcal M}\bigl(\bm{x}^{(i)}-\bm{H}_{\bm{x}^{(i)}}^{-1}\bm{g}_{\bm{x}^{(i)}}\bigr)\label{Newton}
\end{equation}
from a given initial value $\bm{x}^{(0)}\in\mathcal M$ until $(\bm{x}^{(i)})_{i\in \mathbb N}$ converges.
If $\bm{x}^{(0)}$ is sufficiently close to a stationary point $\bm{x}^{\star}\in\mathcal M$ of $h$ (i.e., $\bm{g}_{\bm{x}^{\star}}=\bm{0}$), then $(\bm{x}^{(i)})_{i\in \mathbb N}$ generated by \eqref{Newton} converges to $\bm{x}^{\star}$ at least quadratically \cite[Theorem 6.3.2]{Absil-Mahony-Sepulchre-07}. 
Since a solution to the problem in \eqref{opt_on_M} is also a stationary point of $h$ on $\mathcal M$, the problem is quickly solved if we can give a proper $\bm{x}^{(0)}$.

\section{Tight Minimum-Sidelobe Window}

We propose a class of tight windows that minimize the sidelobe energy.
It is defined as a solution to the \textit{spectral concentration problem} under the constraint of tightness.

\subsection{Spectral Concentration Problem and Slepian Window}

Define the \textit{discrete-time Fourier transform} of $\bm{w}$ as%
\footnote{The sampling interval is set to $1$, but it can be changed to any length.}
\begin{equation}
\widehat{w}(f)=\sum_{l=0}^{K-1}w[l]\,\mathrm{e}^{-2\pi\mathbbm{i} fl}\quad\mbox{for }f\in[-1/2,1/2)\mbox{,}\label{DTFT}
\end{equation}
and let $p\in(0,1)$ be a given proportion of the \textit{mainlobe} to the entire frequency.
To minimize the sidelobe energy of $\widehat{w}$, the following spectral concentration problem is considered:
\begin{equation}
\mathop{\mathrm{maximize}}_{\bm{w}\,\in\,\mathbb R^{K}\setminus\{\bm{0}\}}\,\frac{\int_{-p /2}^{p /2}|\widehat{w}(f)|^2\,\mathrm{d}f}{\int_{-1/2}^{1/2}|\widehat{w}(f)|^2\,\mathrm{d}f}\mbox{.}\label{scp}
\end{equation}
Note that this maximization of the mainlobe energy is equivalent to minimization of the sidelobe energy.
The numerator of the cost function of \eqref{scp} can be expressed as
\begin{align*}
&\int_{-p /2}^{p /2}|\widehat{w}(f)|^2\,\mathrm{d}f=\sum_{l=0}^{K-1}\sum_{l'=0}^{K-1}w[l]w[l']\int_{-p /2}^{p /2}\mathrm{e}^{-2\pi \mathbbm{i} f(l-l')}\,\mathrm{d}f\nonumber\\
&=p \sum_{l=0}^{K-1}\sum_{l'=0}^{K-1}w[l]w[l']\,\mathrm{sinc}(p(l-l'))=\bm{w}^{\mathrm{T}}\bm{Q}_{p}\bm{w}\mbox{,}
\end{align*}
where $\bm{Q}_{p}:=(p\,\mathrm{sinc}(p(l-l')))=(\frac{\sin(\pi p (l-l'))}{\pi (l-l')})\in\mathbb R^{K\times K}$ is a positive-definite symmetric matrix.
For $p=1$, we have
\[
\int_{-1 /2}^{1/2}|\widehat{w}(f)|^2\,\mathrm{d}f=\bm{w}^{\mathrm{T}}\bm{Q}_{1}\bm{w}=\bm{w}^{\mathrm{T}}\bm{I}_{K}\bm{w}=\bm{w}^{\mathrm{T}}\bm{w}\mbox{.}
\]
Thus, the problem in \eqref{scp} maximizes the \textit{Rayleigh quotient},
\begin{equation}
\mathop{\mathrm{maximize}}_{\bm{w}\,\in\,\mathbb R^{K}\setminus\{\bm{0}\}}\,\frac{\bm{w}^{\mathrm T}\bm{Q}_{p}\bm{w}}{\bm{w}^{\mathrm T}\bm{w}}\mbox{,}\label{Rayleigh quotient}
\end{equation}
which is equivalent, by letting $\bm{w}^{\mathrm{T}}\bm{w}=\lVert\bm{w}\rVert_{2}^{2}=1$, to
\begin{equation}
\mathop{\mathrm{maximize}}_{\bm{w}\,\in\, \mathcal S_{1}^{K-1}}\,\bm{w}^{\mathrm T}\bm{Q}_{p}\bm{w}\mbox{.}\label{scp_sphere}
\end{equation}
The solution $\bm{w}_{\mathcal S,p}$ to the problems in \eqref{Rayleigh quotient} and \eqref{scp_sphere} is the \textit{first principal eigenvector} of $\bm{Q}_{p}$ and is symmetric, i.e., $w_{\mathcal S,p}[l]=w_{\mathcal S,p}[K-l-1]$ for $l=0,1,\ldots,K-1$ \cite{Slepian1978}.
This minimum-sidelobe window $\bm{w}_{\mathcal S,p}$ is called the \textit{Slepian window}.

\subsection{Proposed Tight Minimum-Sidelobe Window}

Although the Slepian window is optimal in terms of sidelobe energy, its canonical tight window has poor energy concentration as will be illustrated in Section~\ref{sec:exp}.
This is because the projection onto the set of tight windows in \eqref{eq:tightWin} does not take the spectral characteristics into account.
Thus, we propose to find a minimum-sidelobe window within all tight windows.

Here, we assume $K/a=:J\in\mathbb N$ for simplicity,%
\footnote{If $K/a$ is not an integer, subvectors $\bm{w}_l$ of different lengths are generated, which makes the notation of the definition of $\mathcal M$ in \eqref{oblique} more complicated.}
but this assumption can be removed.
From \eqref{frame_operator} and its periodicity with period $a$, the condition for the tightness of $\bm{w}$ is given by
\[
M\sum_{n=0}^{J-1}|w[l+an]|^2=M\lVert \bm{w}_{l}\rVert_{2}^{2}=\lambda\quad\mbox{(}l=0,1,\ldots,a-1\mbox{),}
\]
with $\lambda>0$ and $\bm{w}_{l}:=(w[l],w[l+a],\ldots,w[l+a(J-1)])^{\mathrm T}\in\mathbb R^{J}$ ($l=0,1,\ldots,a-1$).
Without loss of generality, we set $\lambda=M/a$.
Then, the proposed window is formulated as a solution to the following constrained optimization problem:
\begin{equation}
\mathop{\mathrm{maximize}}_{\bm{w}\,\in\, \mathcal S_{1}^{K-1}}\,\bm{w}^{\mathrm T}\bm{Q}_{p}\bm{w}\quad\mbox{subject to }\forall l\  \lVert \bm{w}_{l}\rVert_{2}^{2}=\frac{1}{a}\mbox{,}\label{proposed_problem}
\end{equation}
which is the problem in \eqref{scp_sphere} with the tightness constraint.
To compute its solution, we derive a fast algorithm as follows.

\subsection{Riemannian Newton's Method for the Problem in \eqref{proposed_problem}}
To derive the fast algorithm, the problem in \eqref{proposed_problem} is reformulated.
First, we sort the components of $\bm{w}$ as $\widetilde{\bm{w}}:=(\bm{w}^{\mathrm T}_{0},\bm{w}^{\mathrm T}_{1},\ldots,\bm{w}^{\mathrm T}_{a-1})^{\mathrm T}\in\mathbb R^{K}$ and accordingly sort $\bm{Q}_{p}$ to construct a positive-definite symmetric matrix $\widetilde{\bm{Q}}_{p}\in\mathbb R^{K\times K}$ satisfying $\bm{w}^{\mathrm T}\bm{Q}_{p}\bm{w}=\widetilde{\bm{w}}^{\mathrm T}\widetilde{\bm{Q}}_{p}\widetilde{\bm{w}}$.
Next, we express the tightness constraint $\lVert \bm{w}_{l}\rVert_{2}^{2}=1/a$ ($\Leftrightarrow\lVert \bm{w}_{l}\rVert_{2}=1/\sqrt{a}$) by a set
\begin{align}
\mathcal M:={}&\bigl\{(\bm{w}^{\mathrm T}_{0},\bm{w}^{\mathrm T}_{1},\ldots,\bm{w}^{\mathrm T}_{a-1})^{\mathrm T}\in\mathbb R^{K}\,\bigl|\,\forall l\ \bm{w}_{l}\in \mathcal S^{J-1}_{1/\sqrt{a}}\bigr\}\nonumber\\
={}&\mathcal S^{J-1}_{1/\sqrt{a}}\times \mathcal S^{J-1}_{1/\sqrt{a}}\times\cdots\times \mathcal S^{J-1}_{1/\sqrt{a}}=\bigl(\mathcal S^{J-1}_{1/\sqrt{a}}\bigr)^{a}\mbox{.}\label{oblique}
\end{align}
This constraint set $\mathcal M$, the \textit{direct product of spheres}, is the \textit{Riemannian product manifold} called \textit{oblique manifold} \cite{boumal2020intromanifolds}.
Since $\widetilde{\bm{w}}\in\mathcal M\Rightarrow\bm{w}\in \mathcal S_{1}^{K-1}$, the problem in (\ref{proposed_problem}) is the same as a minimization problem on the oblique manifold $\mathcal M$:
\begin{equation}
\mathop{\mathrm{minimize}}_{\widetilde{\bm{w}}\,\in\,\mathcal M}\,-\frac{1}{2}\widetilde{\bm{w}}^{\mathrm T}\widetilde{\bm{Q}}_{p}\widetilde{\bm{w}}\mbox{.}\label{scp_manifold}
\end{equation}
The cost function $h(\widetilde{\bm{w}}):=-\frac{1}{2}\widetilde{\bm{w}}^{\mathrm T}\widetilde{\bm{Q}}_{p}\widetilde{\bm{w}}$ is twice continuously differentiable in the Euclidean space, and thus the Riemannian Newton's method in \eqref{Newton} is applicable.

To apply it, we have to derive the metric projection onto the tangent space $\bm{P}_{\widetilde{\bm{w}}}$ in \eqref{gradient} and \eqref{Hessian}, the Riemannian gradient $\bm{g}_{\widetilde{\bm{w}}}$ in \eqref{gradient}, the Riemannian Hessian $\bm{H}_{\widetilde{\bm{w}}}$ in \eqref{Hessian}, and the metric projection onto the oblique manifold $P_{\mathcal M}$ in \eqref{Newton}.

\subsubsection{Metric Projection onto the Tangent Space}

Since $\mathcal M$ is a product manifold as shown in \eqref{oblique}, the tangent space of $\mathcal M$ is the direct product of those of the spheres $S^{J-1}_{1/\sqrt{a}}$:
\begin{equation}
T_{\widetilde{\bm{w}}}\mathcal M=T_{\bm{w}_0}\mathcal S^{J-1}_{1/\sqrt{a}}\times T_{\bm{w}_1}\mathcal S^{J-1}_{1/\sqrt{a}}\times\cdots\times T_{\bm{w}_{a-1}}\mathcal S^{J-1}_{1/\sqrt{a}}\mbox{,}\label{tangent_oblique}
\end{equation}
where $T_{\bm{w}_l}\mathcal S^{J-1}_{1/\sqrt{a}}=\{\bm{v}_{l}\in\mathbb R^{J}\,|\,\bm{w}^{\mathrm{T}}_{l}\bm{v}_{l}=0\}$, and the metric projection $\bm{P}_{\bm{w}_l}:\mathbb R^{J}\to T_{\bm{w}_{l}}\mathcal S^{J-1}_{1/\sqrt{a}}$ onto $T_{\bm{w}_l}\mathcal S^{J-1}_{1/\sqrt{a}}$ is
\begin{equation}
\bm{P}_{\bm{w}_l}=\bm{I}_{J}-a\bm{w}_{l}\bm{w}^{\mathrm T}_{l}\mbox{.}\label{tangent_sphere}
\end{equation}
From \eqref{tangent_oblique} and \eqref{tangent_sphere}, the metric projection $\bm{P}_{\widetilde{\bm{w}}}:\mathbb R^{K}\to T_{\widetilde{\bm{w}}}\mathcal M$ onto $T_{\widetilde{\bm{w}}}\mathcal M$ can be expressed as a block-diagonal matrix
\begin{align*}
\bm{P}_{\widetilde{\bm{w}}}&=\bm{I}_K-a\,\mathrm{diag}\bigl(\bm{w}_{0}\bm{w}^{\mathrm T}_{0},\bm{w}_{1}\bm{w}^{\mathrm T}_{1},\ldots,\bm{w}_{a-1}\bm{w}^{\mathrm T}_{a-1}\bigr)\nonumber\\
&=\bm{I}_{K}-a\bm{W}\bm{W}^{\mathrm T}\mbox{,}
\end{align*}
where $\bm{W}=\mathrm{diag}(\bm{w}_{0},\bm{w}_{1},\ldots,\bm{w}_{a-1})\in\mathbb R^{K\times a}$.

\subsubsection{Riemannian Gradient}

The Riemannian gradient of $h$ at $\widetilde{\bm{w}}\in\mathcal M$ is given by projecting the Euclidean gradient as
\begin{align}
&\bm{g}_{\widetilde{\bm{w}}}=\bm{P}_{\widetilde{\bm{w}}}
\bigl(\nabla h(\widetilde{\bm{w}})\bigr)=-\bigl(\bm{I}_{K}-a\bm{W}\bm{W}^{\mathrm T}\bigr)\widetilde{\bm{Q}}_{p}\widetilde{\bm{w}}\nonumber\\
&=-\Bigl(\widetilde{\bm{Q}}_{p}-a\,\mathrm{diag}\bigl(h_{0}(\widetilde{\bm{w}})\bm{I}_{J},h_{1}(\widetilde{\bm{w}})\bm{I}_{J},\ldots,h_{a-1}(\widetilde{\bm{w}})\bm{I}_{J}\bigr)\Bigr)\widetilde{\bm{w}}\nonumber\\
&=:-\bm{U}_{\widetilde{\bm{w}}}\widetilde{\bm{w}}\mbox{,}\label{pro_gradient}
\end{align}
where $h_{l}(\widetilde{\bm{w}}):=\bm{w}^{\mathrm T}_{l}\bm{q}_{l}$ ($l=0,1,\ldots,a-1$) is a differentiable function computed with $\widetilde{\bm{Q}}_{p}\widetilde{\bm{w}}=(\bm{q}_{0}^{\mathrm T},\bm{q}_{1}^{\mathrm T},\ldots,\bm{q}_{a-1}^{\mathrm T})^{\mathrm T}\in\mathbb R^{K}$.

\subsubsection{Riemannian Hessian}

According to \eqref{Hessian} and \eqref{pro_gradient}, the Riemannian Hessian of $h$ at $\widetilde{\bm{w}}\in\mathcal M$ satisfies
\begin{align}
&\bm{H}_{\widetilde{\bm{w}}}\bm{v} = -\bm{P}_{\widetilde{\bm{w}}}\bm{U}_{\widetilde{\bm{w}}}\bm{v}\nonumber\\
&\qquad+a\bm{P}_{\widetilde{\bm{w}}}\bm{W}\begin{bmatrix}
\nabla h_{0}(\widetilde{\bm{w}}) & \nabla h_{1}(\widetilde{\bm{w}}) & \cdots&\nabla h_{a-1}(\widetilde{\bm{w}})
\end{bmatrix}^{\mathrm{T}}\bm{v}\nonumber\\
&=-\bigl(\bm{I}_{K}-a\bm{W}\bm{W}^{\mathrm T}\bigr)\bm{U}_{\widetilde{\bm{w}}}\bm{v}\nonumber\\
&=-\Bigl(\bm{U}_{\widetilde{\bm{w}}}\bm{v}-a\bm{W}\bm{W}^{\mathrm T}\widetilde{\bm{Q}}_{p}\bm{v}\nonumber\\
&\qquad +a^2\bm{W}\bm{W}^{\mathrm T}\bm{V}\,(h_{0}(\widetilde{\bm{w}}), h_{1}(\widetilde{\bm{w}}), \ldots, h_{a-1}(\widetilde{\bm{w}}))^{\mathrm T}\Bigr)\nonumber\\
&=-\bigr(\bm{U}_{\widetilde{\bm{w}}}-a\bm{W}\bm{W}^{\mathrm T}\widetilde{\bm{Q}}_p\bigl)\bm{v}\label{pro_Hessian}
\end{align}
for all $\bm{v}=(\bm{v}^{\mathrm{T}}_{0},\bm{v}^{\mathrm{T}}_{1},\ldots,\bm{v}^{\mathrm{T}}_{a-1})^{\mathrm{T}}\in T_{\widetilde{\bm{w}}}\mathcal M$, where we exploited two properties of the oblique manifold, $\bm{P}_{\widetilde{\bm{w}}}\bm{W}=\bm{O}_{K\times a}$ and $\bm{W}^{\mathrm{T}}\bm{V}=\bm{O}_{a\times a}$, with $\bm{V}=\mathrm{diag}(\bm{v}_{0},\bm{v}_{1},\ldots,\bm{v}_{a-1})\in\mathbb R^{K\times a}$.
From \eqref{pro_gradient} and \eqref{pro_Hessian}, a stationary point $\bm{v}\in T_{\widetilde{\bm{w}}}\mathcal M$ in \eqref{quad_approx} is
\begin{equation}
\bm{v}=-\bm{H}^{-1}_{\widetilde{\bm{w}}}\bm{g}_{\widetilde{\bm{w}}}=-\bigr(\bm{U}_{\widetilde{\bm{w}}}-a\bm{W}\bm{W}^{\mathrm T}\widetilde{\bm{Q}}_p\bigl)^{-1}\bm{U}_{\widetilde{\bm{w}}}\widetilde{\bm{w}}\mbox{.}\label{solve}
\end{equation}
Using the \textit{matrix inversion lemma}, it can also be expressed as
\[
\bm{v}=-\widetilde{\bm{w}}+\frac{1}{a}\bm{U}_{\widetilde{\bm{w}}}^{-1}\bm{W}\bigl(\bm{W}^{\mathrm T}\bm{U}_{\widetilde{\bm{w}}}^{-1}\bm{W}\bigr)^{-1}\bm{1}\mbox{,}
\]
which guarantees $\bm{v}\in T_{\widetilde{\bm{w}}}\mathcal M$ from $\bm{W}^{\mathrm{T}}\bm{v}=\bm{0}$, where $\bm{1}\in\mathbb R^{a}$ and $\bm{0}\in\mathbb R^{a}$ denote vectors whose components are all ones and zeros, respectively.

\subsubsection{Proposed Algorithm}

The metric projection $P_{\mathcal M}$ onto the oblique manifold $\mathcal M$ can be computed by applying the metric projection $P_{\mathcal S^{J-1}_{1/\sqrt{a}}}$ onto the sphere $S^{J-1}_{1/\sqrt{a}}$ to each subvector $\bm{w}_l$.
Thus, the Riemannian Newton's method in \eqref{Newton} is given by 
\begin{equation}
\bm{w}^{(i+1)}_{l}=P_{\mathcal S^{J-1}_{1/\sqrt{a}}}\bigl(\bm{w}^{(i)}_{l}+ \bm{v}^{(i)}_{l}\bigr)=\frac{\bm{w}^{(i)}_{l}+\bm{v}^{(i)}_{l}}{\sqrt{a}\,\bigl\lVert \bm{w}^{(i)}_{l}+\bm{v}^{(i)}_{l}\bigr\rVert_{2}}\mbox{,}\label{pro_update}
\end{equation}
where $\bm{v}^{(i)}=(\bm{v}^{(i)\mathrm{T}}_{0},\bm{v}^{(i)\mathrm{T}}_{1},\ldots,\bm{v}^{(i)\mathrm{T}}_{a-1})^{\mathrm{T}}\in T_{\widetilde{\bm{w}}^{(i)}}\mathcal M$ is computed by substituting $\widetilde{\bm{w}}^{(i)}$ into \eqref{solve}.
The proposed algorithm is summarized in Fig.\:\ref{alg:prop}.

The initial tight window $\widetilde{\bm{w}}^{(0)}\in\mathcal M$ is important for quickly obtaining a solution.
We recommend two initializations.
One is to use the tight window $\bm{w}^{(0)}=\sqrt{M/a}\bm{S}_{\bm{w}_{\mathcal S,p}}^{-1/2}\bm{w}_{\mathcal S,p}$ obtained from the Slepian window $\bm{w}_{\mathcal S,p}$, i.e., $\widetilde{\bm{w}}^{(0)}=P_{\mathcal M}(\widetilde{\bm{w}}_{\mathcal S,p})$.
Since $\bm{w}_{\mathcal S,p}$ is the solution to the problem in \eqref{scp_sphere}, if $P_{\mathcal M}(\widetilde{\bm{w}}_{\mathcal S,p})\approx\widetilde{\bm{w}}_{\mathcal S,p}$, the solution to the problem in \eqref{proposed_problem} is expected to exist in the close neighborhood of $\bm{w}^{(0)}$.
The other is to use a tight window $\bm{w}_{\mathcal M,p}$ obtained by the proposed algorithm with a different parameter $p$.
Since a solution to the problem in \eqref{proposed_problem} should change continuously with the change in $p$,  if $p'\approx p$, $\bm{w}_{\mathcal M,p}$ is expected to exist in the close neighborhood of $\bm{w}_{\mathcal M,p'}$.

\begin{figure}[t]
\renewcommand{\algorithmicrequire}{\textbf{Input:}}
\small
	\begin{algorithmic}[1]
	\REQUIRE Initial tight window $\bm{w}$, time shift $a$, and proportion $p$.
	\STATE Construct $\bm{Q}_{p}$.
		\STATE Convert $\bm{w}$ and $\bm{Q}_{p}$ to $\widetilde{\bm{w}}$ and $\widetilde{\bm{Q}}_{p}$, respectively.
		\STATE Construct $\bm{U}_{\widetilde{\bm{w}}}$ and $\bm{W}$.
		\STATE $\bm{g}_{\widetilde{\bm{w}}}\leftarrow-\bm{U}_{\widetilde{\bm{w}}}\widetilde{\bm{w}}$
		\STATE $i\leftarrow0$
		\WHILE{$\lVert \bm{g}_{\widetilde{\bm{w}}}\rVert_{2}>\delta$ and $i<i_{\max}$}
		\STATE $\bm{v}\leftarrow\bigr(\bm{U}_{\widetilde{\bm{w}}}-a\bm{W}\bm{W}^{\mathrm T}\widetilde{\bm{Q}}_p\bigl)^{-1}\bm{g}_{\widetilde{\bm{w}}}$
		\STATE $\bm{w}_{l}\leftarrow \frac{\bm{w}_{l}+\bm{v}_{l}}{\sqrt{a}\,\lVert \bm{w}_{l}+\bm{v}_{l}\rVert_{2}}$\quad($l=0,1,\ldots,a-1$)\\[1pt]
		\STATE Construct $\bm{U}_{\widetilde{\bm{w}}}$ and $\bm{W}$.
		\STATE $\bm{g}_{\widetilde{\bm{w}}}\leftarrow-\bm{U}_{\widetilde{\bm{w}}}\widetilde{\bm{w}}$
		\STATE $i\leftarrow i+1$
		\ENDWHILE
		\STATE Convert $\widetilde{\bm{w}}$ to $\bm{w}$.
       \STATE Return $\bm{w}$.
	\end{algorithmic}
	\caption{Proposed algorithm for computing tight minimum-sidelobe windows.}
	\label{alg:prop}
\end{figure}%

\section{Numerical Example}
\label{sec:exp}

In this section, we show some windows designed by varying the mainlobe-width parameter $p\in\{1/K,2/K,\ldots,20/K\}$ while fixing $K=512$ and $a=128$.
For $p=1/K$, the initial value of the proposed algorithm was set to the tight window given from $\bm{w}_{\mathcal S,p}$.
Then, for $p=n/K$ ($n\geq2$), the proposed window for $p=(n-1)/K$ was used as the initial value.
The iteration of the proposed algorithm was terminated
when the norm of Riemannian gradient became smaller than $\delta=10^{-15}$.

\begin{figure}[!t]
\centering
          \includegraphics[width=0.96\columnwidth]{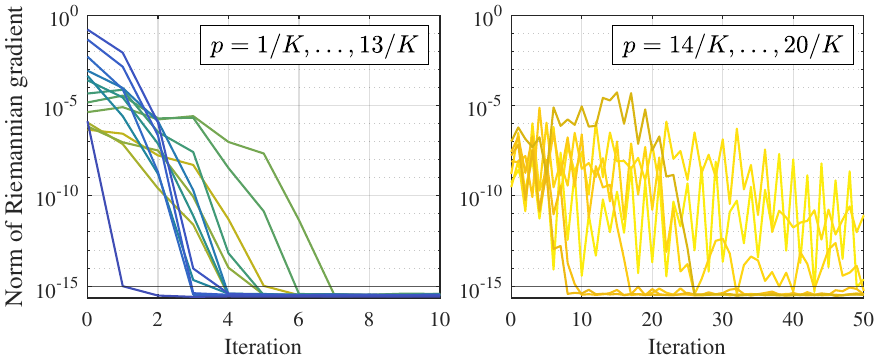}
\caption{The value of the Euclidean norm of the Riemmanian gradient at each iteration $\lVert\bm{g}_{\widetilde{\bm{w}}^{(i)}}\rVert_{2}$.
Since the required numbers of iterations were largely different for some $p$ (see Table~\ref{tab:requiredIter}), the figure was split into two parts: results for $p=1/K,\ldots,13/K$ (left) and those for $p=14/K,\ldots,20/K$ (right).
The color represents $p$, where $1/K$ is dark blue, and $20/K$ is yellow.}
\label{fig:gradIter}
\end{figure}

The norm of Riemannian gradient at each iteration is shown in Fig.\:\ref{fig:gradIter}, where each line corresponds to one of the mainlobe-width parameter $p$ (see Table~\ref{tab:requiredIter}), and the stopping criterion $10^{-15}$ is indicated by the horizontal line.
The numbers of iterations necessary to meet the stopping criterion are summarized in Table~\ref{tab:requiredIter}.
The proposed Newton's algorithm was able to rapidly obtain a solution for $p\leq 13/K$.
The algorithm required more iterations for $p\geq 14/K$, but it was still fast except for $p=19/K$.
The instability for higher $p$ should be because of the numerical ill-conditioning of $\widetilde{\bm{Q}}_p$ \cite{Slepian1978,Verma1996}.%
\footnote{Even though we used an accurate method \cite{Rump2013} for computing the matrix inverse in the 7th line of Fig.\:\ref{alg:prop}, the instability occurred as shown in Fig.\:\ref{fig:gradIter}.}
Note that the proposed Newton's method for $a =1$ coincides with the \textit{Rayleigh quotient iteration} which converges to an eigenvector cubically \cite{Absil-Mahony-Sepulchre-07}.
That is, the proposed algorithm can be considered as its generalized version for the oblique manifold $\mathcal{M}$, and hence the rapid convergence is expected.

\begin{table}[t]
 \caption{Number of iterations required to meet the stopping criterion.}
 \label{tab:requiredIter}
 \centering
  \begin{tabular}{|c|c|}
   \hline
   $p$ &  $i$ \\
   \hline
   \textcolor[rgb]{0.2521,0.3010,0.7096}{$1/K$} & $2$\\ 
   \textcolor[rgb]{0.2386,0.3333,0.7601}{$2/K$} & $4$\\ 
   \textcolor[rgb]{0.2208,0.3723,0.7859}{$3/K$} & $3$\\
   \textcolor[rgb]{0.1960,0.4200,0.7724}{$4/K$} & $3$\\
   \textcolor[rgb]{0.1499,0.4769,0.6995}{$5/K$} & $4$\\
   \hline
  \end{tabular}
  \begin{tabular}{|c|c|}
   \hline
   $p$ &  $i$ \\
   \hline
   \textcolor[rgb]{0.1349,0.5263,0.6176}{$6/K$} & $4$\\
   \textcolor[rgb]{0.1761,0.5677,0.5398}{$7/K$} & $4$\\
   \textcolor[rgb]{0.2530,0.6027,0.4651}{$8/K$} & $5$\\
   \textcolor[rgb]{0.3486,0.6316,0.3923}{$9/K$} & $6$\\
   \textcolor[rgb]{0.4570,0.6541,0.3198}{$10/K$} & $7$\\
  \hline
  \end{tabular}
  \begin{tabular}{|c|c|}
   \hline
   $p$ &  $i$ \\
   \hline
   \textcolor[rgb]{0.5581,0.6733,0.2481}{$11/K$} & $5$\\
   \textcolor[rgb]{0.6533,0.6901,0.1770}{$12/K$} & $4$\\
   \textcolor[rgb]{0.7493,0.7033,0.1088}{$13/K$} & $6$\\
   \textcolor[rgb]{0.8512,0.7110,0.0707}{$14/K$} & $26$\\
   \textcolor[rgb]{0.9280,0.7255,0.0774}{$15/K$} & $8$\\
  \hline
  \end{tabular}
  \begin{tabular}{|c|c|}
   \hline
   $p$ &  $i$ \\
   \hline
   \textcolor[rgb]{0.9675,0.7550,0.0788}{$16/K$} & $10$\\ 
   \textcolor[rgb]{0.9886,0.7924,0.0766}{$17/K$} & $17$\\
   \textcolor[rgb]{0.9978,0.8345,0.0718}{$18/K$} & $26$\\
   \textcolor[rgb]{0.9978,0.8795,0.0643}{$19/K$} & $266$\\
   \textcolor[rgb]{0.9902,0.9265,0.0536}{$20/K$} & $43$\\
  \hline
  \end{tabular}
\end{table}

\begin{figure}[!t]
\centering
          \includegraphics[width=0.96\columnwidth]{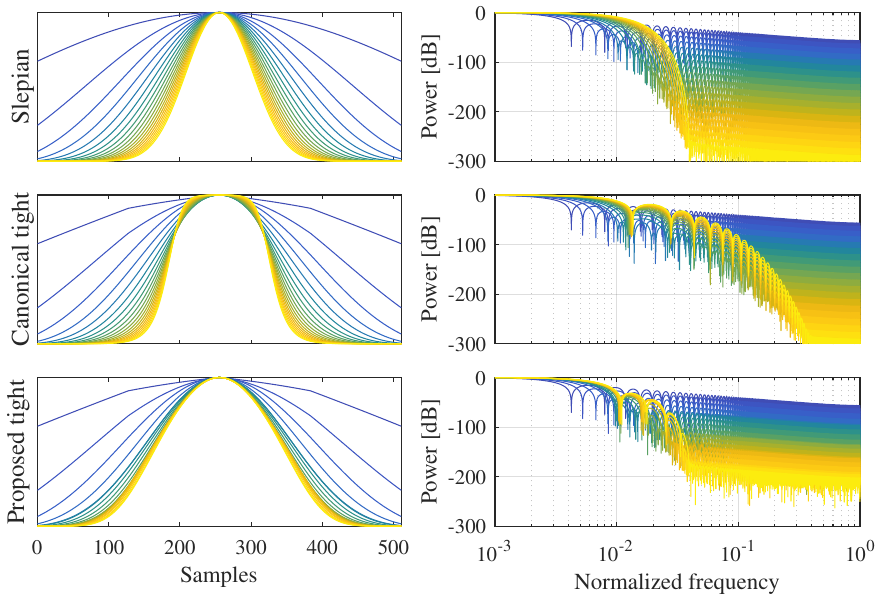}
\caption{Comparison of the conventional and proposed windows ($K=512$, $a=128$, $p=1/K,\ldots,20/K$).
From top to bottom, the Slepian window, its canonical tight window, and the proposed tight window are shown.
The color represents the mainlobe-width parameter $p$, where $1/K$ is dark blue, and $20/K$ is yellow.
All lines are peak normalized, and the frequency axes are normalized so that the Nyquist frequency becomes $10^0$.
}
\label{fig:widows}
\end{figure}

Shapes%
\footnote{We empirically found that all windows were nonnegative and symmetric.}
and spectra of the obtained windows are shown in Fig.\:\ref{fig:widows}.
The Slepian window (upper row) becomes narrower as $p$ increases, which corresponds to increase of the distance from the set of tight windows.
Therefore, its canonical tight window (middle row) is more different for larger $p$.
This modification broke the energy-concentration property of the Slepian window as in the middle right figure.
In contrast, the proposed window (bottom row) was able to narrow the mainlobe as the Slepian window while maintaining the tightness.

\section{Conclusion}

In this paper, we proposed a class of tight windows that minimize the sidelobe energy.
Those windows are characterized by solutions to the energy minimization problem on the oblique manifold.
We also proposed the Riemannian Newton's method for rapidly computing them.
Improvement of the stability of the algorithm for a large window-width parameter $p$ as well as applications of the proposed windows to DGT-domain signal processing are left as future works.


%





\ifCLASSOPTIONcaptionsoff
  \newpage
\fi



\bibliographystyle{IEEEtran}
\bibliography{main}
\end{document}